# Frequency comb from a microresonator with engineered spectrum

**Ivan S. Grudinin,[1,*] Lukas Baumgartel,[1] and Nan Yu[1]**

[1]*Jet Propulsion Laboratory, California Institute of Technology, 4800 Oak Grove Drive, Pasadena, California 91109-8099 USA*
[*]*grudinin@jpl.nasa.gov*

**Abstract:** We demonstrate that by varying the ratio between the linewidth and dispersion of a whispering gallery mode resonator we are able to control the number $N$ of free spectral ranges separating the first generated comb sidebands from the pump. We observed combs with $N$=19 and $N$=1. For the comb with $N$=1 we have achieved a span of over 200 nm using a 0.4 mm $MgF_2$ resonator with a pump of 50 mW, which is a factor of 10 lower than previously reported.

## 1. Introduction

Frequency combs are important for many applications including precise spectroscopy, optical waveform synthesis, environmental sensing, metrology and optical clocks [1]. Recently a new class of combs based on whispering gallery mode (WGM) and ring resonators has been discovered [2-5]. Evidence of octave spanning comb spectra, important for stabilization of repetition rate by self-referencing technique, has recently been reported with silica microtoroids [6] and silicon nitride microrings [7].

While microcavity combs are compact, the efficiency of generation is still low. Microcavity combs require hundreds of milliwatts of optical pump power to achieve relatively wide spectral coverage [6-9]. In addition, the existing microcavity combs may work in two distinct regimes defined by the number $N$ of cavity free spectral ranges (FSR) that separates the first oscillating sideband from the pump. Combs with $N=1$ have stable and narrow microwave beatnotes, which is necessary for many metrology and spectroscopy applications. In the case of $N>1$ the spectrum consists of many overlapping combs, leading to broad microwave beatnotes and conversely an unstable repetition rate [8]. Another problem is the mode crossings caused by a very rich spectrum of conventional optical WGM resonators. The crossings locally disturb cavity spectrum and adversely affect multiplication of comb lines [10, 11]. Producing the sharp-edge cavities is possible to reduce the number of mode families, but the number of high $Q$ modes is still too large in such resonators.

Careful design of the resonator spectrum is required to achieve efficient comb generation in the coherent regime, with low pump power and broad frequency span. We have developed a cavity engineering approach to make a resonator having only one pair of fundamental TE and TM WG modes with high $Q$ and good coupling efficiencies. In order to preserve a high $Q$ of fundamental modes the resonator is allowed to support a small set of non-fundamental modes but with lower $Q$ factor and coupling efficiency. Such a cavity combines the simplicity of the spectrum found in single mode cavities [12] and high optical $Q$ achievable in multimode cavities used today in many laboratories [9, 13-15]. We here present the comb obtained with such resonator for the first time. We show that by controlling the ratio of cavity linewidth to dispersion one can control the separation N of initial comb sidebands.

## 2. Cavity dispersion and linewidth

Comb sideband spacing N in the units of cavity FSR is determined by a combined effect of the cavity dispersion and nonlinear mode shifts due to self- and cross- phase modulation. The relevant figure of merit for the design of low phase noise WGM Kerr combs is the ratio of the cavity linewidth δf to the dispersion parameter $D_2$. In the case of on-resonance pumping it was shown that $N = \sqrt{\delta f / D_2}$ [8]. Here $D_2 = (f_{m+1}-f_m)-(f_m-f_{m-1}) = f_{m+1}+f_{m-1}-2f_m$, where $f_m$ is the frequency of WGMs TE$_{mm1}$ or TM$_{mm1}$, using notations of modes in a spherical cavity [17].

In Fig. 1 we present the dependence of $D_2$ on the radius $R$ of an MgF$_2$ resonator for the TM mode families. In calculating the dependence, we used an approximation for the mode frequencies of a spheroid with major axis $a$ and minor axis $b$ [16,17]. The iterative approach with Sellmeier equations for refractive index of MgF$_2$ [18] was used to compute $D_2(R)$.

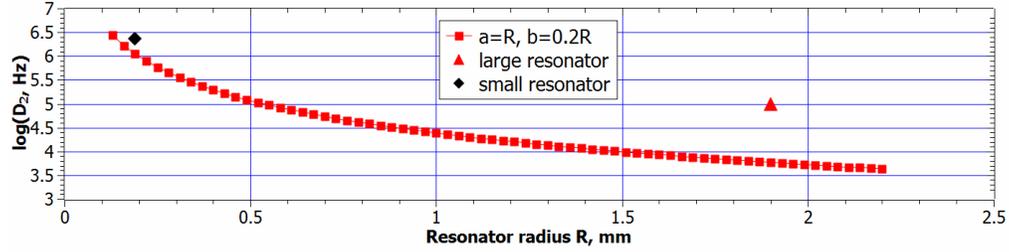

Fig. 1. Dispersion parameter $D_2$ as a function of the radius of a spheroidal z-cut $MgF_2$ resonator for TM modes. Also marked are the two resonators described in Section 3. Parameters for the smaller resonator are: $R$=0.2 mm, $\delta f$=2.3 MHz (loaded linewidth), $D_2$=1 MHz. Parameters for the larger resonator are: $R$=1.9 mm, $\delta f$=0.1 MHz (loaded linewidth), $D_2$=0.0059 MHz.

Fig. 1shows that the condition $\delta f \approx D_2$ is hard to achieve for large resonators, as $D_2$ becomes small and ultrahigh optical $Q$ is required. While a linewidth of 5 kHz has been observed in $CaF_2$ [19], the best unloaded linewidth of $MgF_2$ resonators achieved by us is around 50 KHz ($Q$=3.8×10$^9$, $\lambda$=1.56 nm.) Other groups have reported similar or lower $Q$ factors [9,13]. One also observes that a cavity with a radius of 0.2 mm will produce a comb with $N$=1 at resonance linewidth of around 1 MHz.

## 3. Experimental results

We excited WGMs with a continuous wave (cw) tunable Koheras Adjustik laser having 5 kHz linewidth near $\lambda$=1560 nm. A commercial EDFA was used to achieve pump power of up to 110 mW. Yokogawa AQ6319 optical spectrum analyzer (OSA) was used to detect the comb lines. The schematics of experimental setup are shown on Fig. 2.

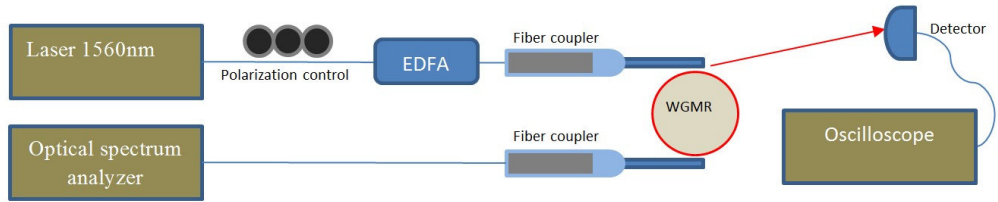

Fig. 2. Schematics of the experimental setup.

Angle-polished [20] fiber couplers are positioned with the XYZ stages and the coupler-resonator gaps are controlled with piezo elements. The couplers are made of SM-28 single mode fiber with core diameter of around 9 micrometers.

*3.1. Large resonator*

To test the predictions described in the previous section we fabricated two z-cut $MgF_2$ resonators. We first tested a resonator made with 3.8 mm diameter, having a sharp edge of around 60 micrometer in radius. The best-coupled mode near 1560 nm was pumped and the laser detuning was adjusted to increase the intracavity power gradually. The first parametric sidebands appeared at $N$=19 cavity FSR (19×18.2 GHz; 2.84 nm). Subsequent comb lines filled up the spectrum as power increased (Fig. 3).

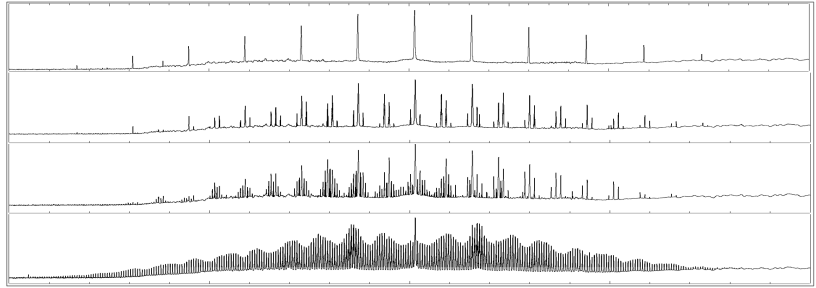

Fig. 3. The comb generated in a 3.8mm MgF$_2$ resonator starts at 19 cavity FSRs around the pump at 1560.29 nm and gradually fills the spectrum with secondary comb lines as the intracavity power is increased. Horizontal span of the plot is 40 nm.

This regime of comb generation produces incoherent combs, where many closely spaced sets of frequency combs exist and the resulting beatnote is broadened by the relative offsets of these combs [8]. When the resonator was pumped with a maximum power of 55 mW at loaded $Q=1.9\times10^9$ the comb contained only lines spaced by 1 FSR and is shown on Fig. 4.

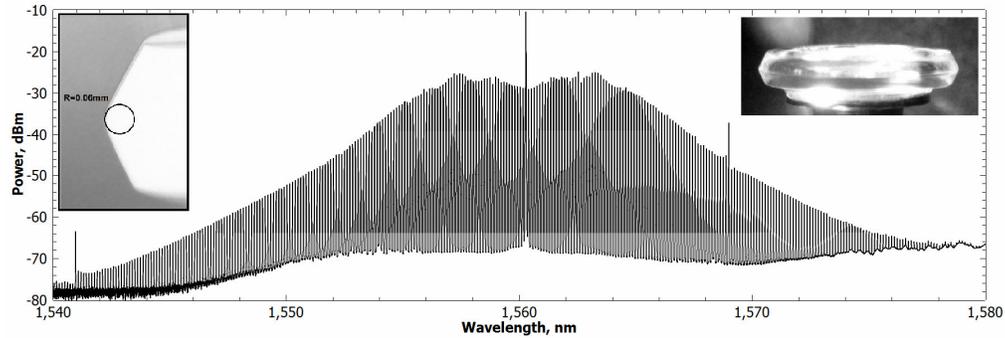

Fig. 4. Comb from a Z-cut MgF$_2$ resonator 3.8 mm in diameter with a sharp edge geometry (60 micrometers radius), loaded linewidth 100 KHz ($Q=1.9\times10^9$), pumped 110 mW at 50% coupling efficiency. 300 lines spaced by 18.204 GHz are present spanning over 40 nm. The two visible envelope irregularities (1541 nm, 1569 nm) could be explained by modal crossings.

*3.2 Small resonator with engineered spectrum*

Following the approach first introduced in [21, 12] we designed and fabricated a second resonator as a ridge waveguide on a cylinder made of a Z-cut MgF$_2$. This resonator was expected to have only one pair of high-Q TE and TM modes. Other higher order modes are reduced in Q and coupling efficiency due to the resonator geometry. To make this resonator we first used a computer-controlled diamond turning process to fabricate a trapezoidal ridge waveguide with dimensions exceeding those required for single mode operation [12]. The polishing step produced a nearly Lorentzian-shaped waveguide shown as the inset in Fig. 6. This nearly single mode resonator geometry is a good alternative to the strictly single mode geometry with a smaller ridge where the WGMs coupling to cylinder modes makes it difficult to have both high Q and small radius. Further analysis is required to explore possible resonator designs.

To model the experimentally excited WGMs in this resonator we used a finite element method (FEM) for numerically solving the vector wave equation [21, 22]. We found all the modes supported in the model resonator within a 4 nm window around the pump wavelength of 1560.3 nm. The lowest order modes are shown as an inset in Fig. 6. The model resonator is supported by a cylinder 391 micrometers in diameter, which was adjusted to approximately match the TE-mode FSR of 172.44 GHz observed in the real resonator. The computed fundamental TE and TM modes matched well with the measurements. However, the higher order modes did not fit well, only producing a similar pattern, which is explained by high sensitivity of spectrum to resonator's shape variation. We found that the comb was generated in a $TE_{1101,1101,1}$ mode (TE1 in Fig. 6). Another observed well-coupled mode was a $TM_{1091,1091,1}$ (TM1 in Fig. 6). The higher measured Q and weaker coupling for this mode is explained by the deeper location of the field of TM modes.

As is evident from modeling, the higher order WGMs no longer fit into the protrusion that is only 6 micrometers high. For the higher order modes, the optical energy is concentrates at the maxima farthest from the equatorial plane. This leads to leakage of optical energy into the cylinder and contributes to weaker coupling of these modes. This helps to explain why only a few modes were observed in this cavity.

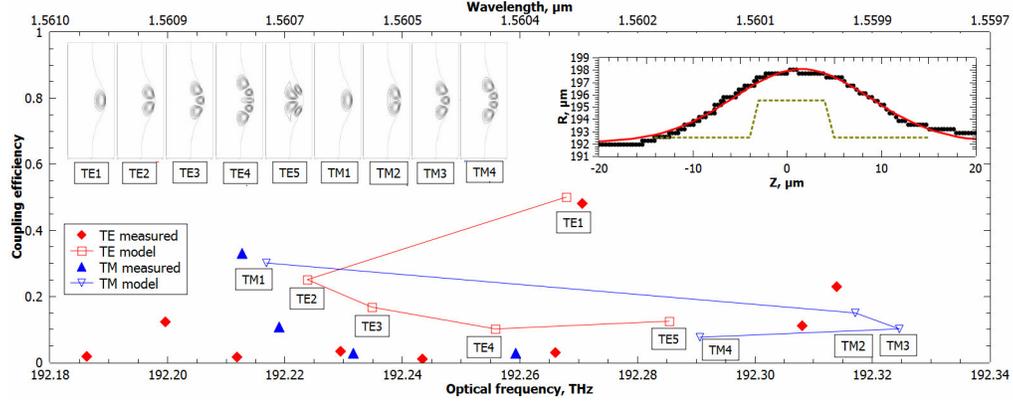

Fig.6. Experimentally measured modes and results of FEM. Coupling efficiency of the computed modes $TE_{lmq}, TM_{lmq}$ is arbitrary set as the inverse product of *(l-m)* and *q*. The two experimentally excited modes were identified as TE1 ($TE_{1101,1101,1}$, $\lambda$=1560.3 nm, intrinsic $Q$=1.92×10$^8$, loaded $Q$=8.3×10$^7$ coupling efficiency 47%), and TM1 ($TM_{1091,1091,1}$, $\lambda$=1560.8 nm, intrinsic $Q$=5.1×10$^8$, loaded $Q$=1.24×10$^8$, coupling efficiency 36%). While TM1 had higher intrinsic *Q*, only 3 pairs of comb sidebands were observed due to lower coupling efficiency and lower loaded *Q*. Top left insets show computed density of electromagnetic energy in WG modes. Top right inset shows the measured profile and a Gaussian approximation of the resonator's shape. Dashed line shows the single mode geometry.

One of the combs observed in the ridge-waveguide resonator is presented in Fig. 7 and 8. In contrast to *N*=19 comb (Fig 3), we now observe an *N*=1 comb (Fig. 7). As we changed the laser detuning to increase the power in the cavity the number of sidebands grew from 1 to 3. Instead of the transition from 3 to 4, a sudden transition from 3 to around 100 sidebands was observed.

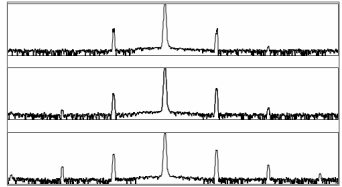

Fig.7. Comb observed at constant pump power with different laser detuning from the cavity resonance. Smooth transition from 1 to 3 sidebands is followed by a jump to a comb shown on Fig. 8. The horizontal span is 9 nm.

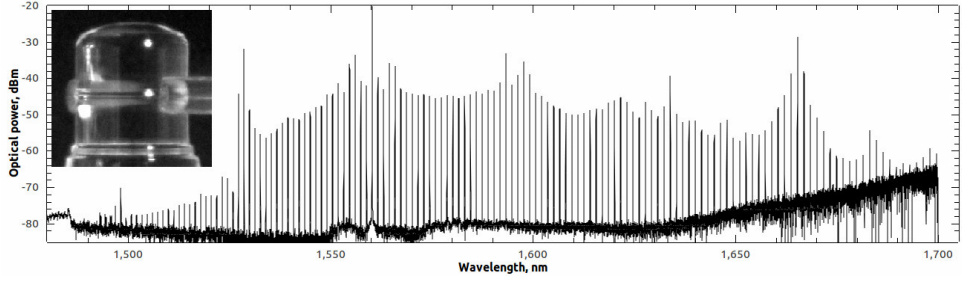

Fig. 8. Frequency comb observed in a resonator with an engineered spectrum. The $TE_{1101,1101,1}$ mode near 1560.3 nm (loaded $Q=8.4\times10^7$, intrinsic $Q=2\times10^8$) was pumped. Resonator diameter is 403 µm. Over a hundred of comb lines spanning more than 200 nm (23.5 THz), limited by OSA range, are observed with only 50 mW of optical pump power.

This behavior could be an evidence of hard comb excitation [24]. The envelope of the generated comb could become more or less regular compared to Fig. 8 depending on coupling conditions. $MgF_2$ is known to have 4 Raman active phonon modes [25], of which the 410 cm$^{-1}$ mode is the strongest. With a pump at 1560.3 nm the Raman Stokes wavelength is expected to be 1667 nm. Raman gain might explain the asymmetry of the comb envelope visible in Fig. 8. However, the sharp drop below 1530 nm suggests that mode crossing could still be involved.

## 6. Discussion

Comparing our results to previously reported combs in $MgF_2$ (Table 1), one can see that the engineered cavity reported in this study demonstrates the best combination of pump power and comb span and also the largest comb repetition rate obtained with a $MgF_2$ resonator.

Table 1. Parameters of various MgF2 microresonator-based frequency combs

| Reference | FSR, GHz (diameter, µm) | Optical Q factor near $\lambda$=1.55 µm | Pump, mW | Pump $\lambda$, µm | Comb span, nm |
|---|---|---|---|---|---|
| [9] | 107 (700) | >$10^9$ | 600 | 2.45 | ~200 |
| [8] | 68 (1000) | ~$2\times10^8$ | 500 | 1.56 | ~300 |
| [13] | 34.67 (2000) | $10^9$ | 2 | 1.543 | ~20 |
| This work | 172.44 (403) | ~$2\times10^8$ | 50 | 1.56 | >200 |

A low-power comb presented in ref. [13] spans 20 nm with pump power of 2 mW. We found, however, that simple increase of the pump power doesn't always produce a broader comb. While we often observed 20 nm spanning combs with a few milliwatt of the pump power in large resonators, further increase of the pump power to 50 mW only produced a comb span of 40 nm as shown in Fig. 4. It should also be noted that during fabrication of the engineered resonator, its $Q$ was initially lower, and we observed comb generation starting with N=2. Non-optimal angle and core diameter of the fiber coupler explains relatively low (50%) coupling efficiencies observed in the experiments.

## 7. Conclusion

We demonstrate that by changing the ratio of the cavity linewidth to its dispersion one can control the number $N$ of the cavity FSRs separating the first comb sidebands from the pump. We present a comb spanning over 200 nm with a pump power 10 times lower than previously reported. We expect that careful engineering of the cavity spectrum is a path to low threshold, high efficiency octave spanning microcavity comb generation that will enable compact optical clocks and other applications.

**Acknowledgments**

The research described in this paper was carried out at the Jet Propulsion Laboratory, California Institute of Technology, under a contract with the NASA. I.S.G thanks M. L. Gorodetsky, A. B. Matsko and D. Strekalov for helpful discussions.